\documentclass[aps,prb,twocolumn,superscriptaddress,showpacs,amsfonts]{revtex4-1}

\usepackage{epsfig}
\usepackage{graphicx}
\usepackage{amsmath,ulem}
\usepackage{amssymb}
\usepackage{color}
\usepackage{bm}

\newcommand{\None}{N$_0=1$~}
\newcommand{\Ntwo}{N$_0=2$~}

\newcommand{\ua}{\uparrow}
\newcommand{\da}{\downarrow}

\begin{document}

\title{Control of Edge Currents at a Ferromagnet - Triplet Superconductor Interface\\ by Multiple Helical Majorana Modes}
\author{Damien~Terrade}
\email[Electronic address: ]{d.terrade@fkf.mpg.de}
\affiliation{Max Planck Institute for Solid State Research,
D-70569 Stuttgart, Germany}
\author{Dirk~Manske}
\affiliation{Max Planck Institute for Solid State Research,
D-70569 Stuttgart, Germany}
\author{Mario~Cuoco}
\affiliation{CNR-SPIN, Fisciano (Salerno), Italy\\ and Dipartimento
di Fisica "E.R. Caianiello", Universit\`{a} di Salerno, Italy}
\date{\today}

\begin{abstract}We study the spin and charge currents flowing at the interface of an itinerant ferromagnet with a topological spin-triplet superconductor having different number of time-reversal-invariant Majorana helical modes. Depending on the number of helical modes, the capacity of carrying spin and charge currents is shown to be directly related to the amplitude and orientation of the ferromagnetic magnetization with respect to the superconducting $\vec{d}$-vector. Differently from the one-helical mode spin-triplet superconductor, we find that the presence of a finite amount of electronic hybridization with the two pairs of Majorana helical modes leads to nonvanishing charge current independently of the ferromagnetic exchange. The competition between the two pairs of Majorana helical modes remarkably yields a spin-current response that is almost constant in the range of weak to intermediate ferromagnetism. The behavior of the spin current is tightly linked to the direction of the spin-polarization in the ferromagnet and tends to be flatten for a magnetization that is coplanar to the spin-triplet $\vec{d}$-vector independently of the number of helical modes.
\end{abstract}

\maketitle


\section{Introduction}

The past few decades have been marked by a growing interest in the study of the interplay between superconductivity and ferromagnetism in heterostructures both for the potential
application in the field of spintronics~\cite{Takahashi2012} and because of the underlying fundamental
physics.~\cite{Zutic04,Buzdin05,Bergeret05,Eschrig08}
The physical properties of ferromagnet (FM)-superconductor (SC)
heterostructures are strongly dependent on the interface and on the nature of the electronic states that are formed within both regions. Proximity effects between magnetism and superconductivity at the cross-talk region close to the interface  can alter their characters in such a way to influence the overall spin- and charge-response of the heterostructure.
 
A special position in the variety of the electronic states is
taken by the gapless modes at the boundary of materials whose bulk
is gapped and owe their existence on the global symmetries of the
bulk state without depending on the details of the surface
scattering and other sample-dependent
parameters~\cite{Hasan2010Rev,Qi2010Rev,Moore2010}. 
Simple band
insulators or conventional superconductors do not support robust
low-energy states at the boundary. The topological non-trivial nature of
the bulk state and the bulk-boundary
correspondence theorem are the fundamental aspects that dictate the existence of surface
states
~\cite{Hasan2010Rev,Qi2010Rev,Moore2010,d-wave,Eschrig:PT}.
Topological phases of matter marked by the existence of protected
gapless surface states and fully gapped bulk excitations are at the forefront of 
condensed-matter research.
Recent advancements in this context
led to the realization of topological insulators and
provided indications that also spin liquid and superconductors
can be topologically non trivial.\cite{Hasan2010Rev,Qi2010Rev,Moore2010}
The quantum-Hall state is a prominent example of such topological states in which the Hall
conductance is identified with the topological number as introduced
by Thouless, Kohmoto, Nightingale, an den Nijs
(TKNN).~\cite{Thouless1982} 
Within the superconducting systems, a notable case of superconductor with non-trivial TKNN number is 
the two dimensional ($p$+i$p$)-wave superconductor with
time reversal symmetry breaking,  
which has in Sr$_2$RuO$_4$ its leading candidate.~\cite{Maeno98,Mackenzie2003,Kallin2012}
The rapid growth of interest in this area led to the proposal~\cite{Kane2005,Moore2007,Fu2006,Fu2007,Fukui2007,Sheng2006,Qi2006,
Schnyder2008,Roy2006,Qi2009,Tanaka2012} of an additional class of
topological superconductors with time reversal invariance, referred to as a DIII-symmetry-class superconductor and
classified by the $\mathbb{Z}_2$ topological invariant, which currently is at center of an intense
investigation~\cite{Schnyder2008,Roy2006,Qi2009,Tanaka2012,
Teo2010,Schnyder2010,Beenakker2011,Deng2012, Nakosai2012, Wong2012, Zhang2013, Nakosai2013}.
This type of superconductors can be seen as the time reversal partner of the chiral ones, in a similar 
fashion as the quantum spin Hall systems relate to integer quantum Hall systems.
Differently from chiral superconductors and in analogy with quantum spin Hall systems, those with time reversal invariance can have zero modes that come in pairs, due to Kramers's degeneracy, and can support 
counterpropagating edge states of opposite spins near the boundary that carry a spin current. 
Proposals for the realization of $\mathbb{Z}_2$ time reversal invariant superconductors include spin-singlet 
$s-$, $d-$ and $s\pm$-wave, spin triplet $p$-wave as well as mixed-parity pairing both in bulk and heterostructures 
with proximity to ferromagnet and semiconductors.

\begin{figure}[!t]
\begin{center}
\includegraphics[width=0.5\textwidth]{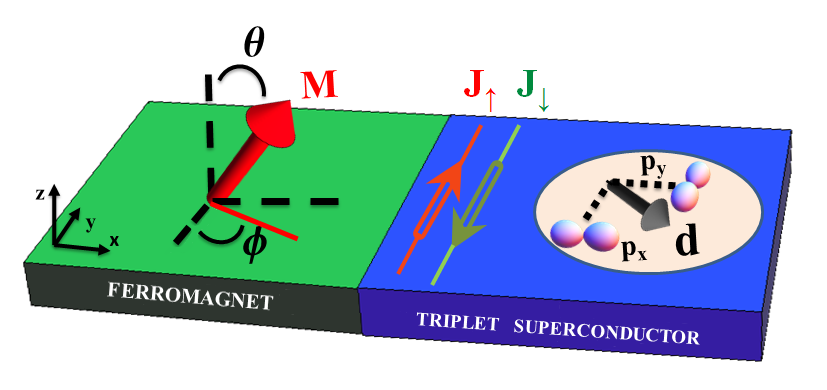}
\end{center}
\caption{(color  online).  Schematic view of the FM-TTSC heterostructure with the
interface perpendicular to the $x$-direction. The right side (blue)
shows the spin-triplet superconductor region, having a $\vec{d}$-vector (thick black arrow) that lies 
within the ($xy$)-plane with orbital components $(d_x,d_y)=(p_y,p_x)$. 
$J_{\uparrow/\downarrow}$ are the spin-polarized currents flowing along the interface
due to the presence of helical edge states in the superconductor.
The left side (red) indicates the ferromagnetic region. The red thick arrow
stands for the magnetization $\vec{M}$  in the interior of the ferromagnet due to an exchange 
field forming relative angles $\theta$ (out-of-plane) and $\phi$ (in-plane) with respect to the $\vec{d}$-vector.}
\label{fig.100.fm.ttsc.interface}
\end{figure}

Among pure time reversal invariant spin-triplet superconductors,
candidate materials would include the $^3$He B phase,\cite{Voll90,Schnyder2008,Roy2006,Qi2009,Chung2009} Cu-doped Bi$_2$,\cite{Fu2010,Hsieh2012,Sasaki2011} 
p-type TlBiTe$_2$, \cite{Yan2010} at the interface of Sr$_2$RuO$_4$ \cite{Tada2009}, and in BC$_3$.\cite{Chen2014}
Remarkably, the search for exotic topological states 
opened novel frontiers that bring the possibility of having topological $p$-wave superconducting phase with time reversal invariance 
in doped Mott insulators \cite{Hyart2012,You2012,Okamoto2013} 
as described by the Kitaev-Heisenberg model.\cite{Jackeli2009, Chaloupka2010}
Due to the rich physical scenario, the study of spin-triplet superconductor interfaced to itinerant ferromagnet has 
relevant implications for getting deeper insight into the nature of topological states and 
on the appealing possibility to achieve new spintronic devices based on the 
coupling between magnetism and superconductivity.
Most of the recent research efforts focused on heterostructure based on chiral or single-component spin-triplet superconductors and
ferromagnetic systems \cite{Hirai2001, Kuboki2004, Yokoyama2007a, Yokoyama2007b, 
Cuoco2008, Brydon2008a, Brydon2008b,Brydon2009,
Gentile2013,Terrade2013} motivated by the potential realization of prototypes using Sr$_2$RuO$_4$ as 
spin-triplet superconductor.\cite{Anwar2014}   
The interface of helical spin triplet superconductors with magnetism is quite promising since, it makes possible the manipulation of the spin currents at the edge of the superconductor with respect to the magnetization inside the ferromagnet, in a similar fashion than in spintronic. Along this direction, the case of heterostructures based on ferromagnet interfaced to non-centrosymmetric superconductors with nontrivial topological states showed the edge current to exhibit dramatic different behaviors underlying the inequivalent topological properties.\cite{SchnyderTimm2013}    

In this paper we investigate the coupling between an itinerant ferromagnet (FM) and a topological
spin-triplet superconductor (TTSC) having different number ($N_0$) of time-reversal-invariant Majorana helical modes (see Fig. \ref{fig.100.fm.ttsc.interface}).
Using topological
invariants it has been recognized that the character and the number of Majorana helical modes are
intimately linked to the topological invariants of the Fermi
surface in the normal state.~\cite{Sato2009} We employ this finding to 
design the electronic structure of the spin-triplet superconductor in order to have
one- and two- pairs of helical modes at the interface with the ferromagnet. The main goal of the analysis is 
to understand how the behavior of spin and charge currents depends on the number of helical modes
in the presence of ferromagnetism. 
This issue is faced by obtaining the spatial profile of the superconducting and magnetic order parameters via a 
self-consistent solution of the Bogoliubov-De Gennes equations for a two-dimensional planar heterostructure and, then,  
the resulting spin- and charge-currents at the interface. 
We find that 
the intensity for the spin and charge currents rescaled to the vacuum interface generally increases with the number of helical modes. 
An important element in the control of the interface current is represented by the degree of mixing 
between the ferromagnetic states and the helical modes. This effective hybridization is linked to the
amplitude and orientation of the spin polarization in the ferromagnet with respect to the superconducting $\vec{d}$-vector as well as 
to the topology of the ferromagnet Fermi surface. Such microscopic aspects 
are shown to play an important role in determining the spin- and charge-current response of the 
heterostructure. 
We show that, differently from the one-helical mode spin-triplet superconductor, the 
presence of a finite amount of electronic hybridization
with the double pairs of Majorana helical modes 
leads to nonvanishing charge current independently on the strength of the ferromagnetic exchange. 
The unequal mixing with the ferromagnetic states of the two Majorana helical modes remarkably yields a spin-current response that is almost
constant for the case of exchanges ranging from weak to intermediate ferromagnetism. Finally, the spin current exhibits trends that are significantly influenced by the orientation of the ferromagnetic magnetization. As a function of the magnetization's strength, we find monotonous or almost flat behaviors for magnetization directions that are perpendicular and coplanar to the spin-triplet $\vec{d}$-vector, independently of the number of helical modes.

The remainder of the paper is organized as follows. We present in Sect.~II the model and the formalism for the ferromagnet- spin triplet superconductor. Sect.~III is devoted to the analysis of the spectral functions close to the 
interface as a function of the magnetization's strength. We analyze in subsections A and B from the Sect.~IV the spatial dependence of the edge currents at the FM-TTSC interface assuming the magnetization in the ferromagnet to be perpendicular and coplanar to the $\vec{d}$-vector, respectively. Finally, the Sect.~V is devoted to the concluding remarks.


\section{Theory}

\subsection{Model and formalism}
\begin{figure}[!t]
\begin{center}
\includegraphics[width=0.5\textwidth]{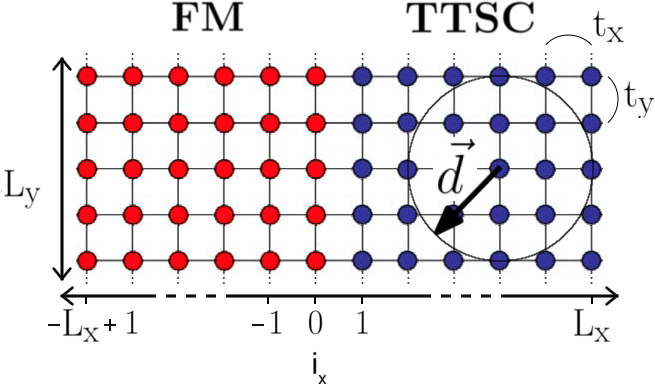}
\end{center}
\caption{(color  online).  View of the $L_x \times L_y$ FM-TTSC lattice with an in-plane $\vec{d}$-vector and where $t_x$ and $t_y$ are the hopping terms in the $x$- and $y$-directions, respectively. The interface between the two regions is situated between the sites $i_x=0$ (FM) $i_x=1$ (TTSC).  }
\label{fig:discrete-lattice}
\end{figure}

We consider a two-dimensional (2D) FM-TTSC heterostructure on a lattice (Fig.~\ref{fig:discrete-lattice}) 
described by a single-band tight-binding model Hamiltonian. 
An attractive  nearest-neighbor interaction is added to yield a spin-triplet pairing with helical 
symmetry in the superconducting side and, following the Stoner model for itinerant magnets, 
an exchange field $h$ is introduced to provide a nonzero spin polarization in the ferromagnetic region. 
The interface is chosen perpendicular to the $x$-direction and the lattice size is $L_x \times L_y$ with periodic boundary 
conditions imposed along the $y$-direction parallel to the interface. Each site of the lattice is indicated by a vector 
$\mathbf{i}\equiv(i_x,i_y)$, with $i_x$ and $i_y$ denoting the site positions in the $x$ and $y$-directions. 
The Hamiltonian is expressed as
\begin{eqnarray}
H&=& -\sum_{\langle \mathbf{i} ,\mathbf{j} \rangle,\,\sigma}
t_{\mathbf{i}\mathbf{j}}(c^{\dagger}_{\mathbf{i}\,\sigma}
c_{\mathbf{j}\,\sigma}+H.c.) -\mu \sum_{\mathbf{i},\sigma}
n_{\mathbf{i}\sigma} \nonumber \\&& - \sum_{\langle \mathbf{i}
,\mathbf{j} \rangle \in \text{TTSC}} V{^{\sigma\sigma{^{'}}}}
n{_{\mathbf{i} \sigma}} n{_{\mathbf{j}\sigma{^{'}}}} -
\sum_{\mathbf{i} \in \text{FM}} \vec{h}\cdot\vec{s}(\mathbf{i}),
\label{eq.hamiltonian}
\end{eqnarray}
\noindent where $c_{\mathbf{i}\,\sigma}$ is the annihilation operator of an electron with spin $\sigma$ at 
the site ${\mathbf{i}}$, $n_{\mathbf{i}\,\sigma} = c^{\dagger}_{\mathbf{i}\,\sigma} c_{\mathbf{i}\,\sigma}$ is the spin-$\sigma$ number 
operator and $t_{\mathbf{i}\mathbf{j}}$ is the hopping amplitude that is nonvanishing only between the nearest neighboring
sites ${\mathbf{i}}$ and ${\mathbf{j}}$. Moreover, $\vec{s}(\mathbf{i}) =\sum_{s,s'}c^{\dagger}_{\mathbf{i}\,s}{\vec{\sigma}}_{s,s'}c_{\mathbf{i}\,s'}$ is the local 
spin density polarization and $V{^{\sigma\sigma{^{'}}}}$ is the pairing coupling between nearest-neighbors. The lattice is divided in two regions, the ferromagnetic side is located at $i_x \leq 0 $ while the superconducting is  
at $i_x > 0$.

Concerning the properties of the ferromagnet, the magnetization is proportional to the exchange field $\vec{h}$ which leads to a splitting of the spin up and spin down energy spectrum and induces a spin-polarisation. Its orientation is given by fixing the angles $\theta$ and $\phi$ with respect to the direction of the $\vec{d}$-vector (see Fig.~\ref{fig.100.fm.ttsc.interface}).
Therefore, the magnetization is coplanar to the $\vec{d}$-vector for $\theta=0$ while it is perpendicular to the $\vec{d}$-vector when $\theta=\pi/2$. 
Moreover, in the case $\theta=0$, the spin-polarization is collinear to the x-component (y-component) of the $\vec{d}$-vector for $\phi=0 (\pi/2)$, respectively. The $\phi$ angular dependence is nontrivial because it couples to zero spin projections having an inequivalent orbital symmetry.
Although the topology of the Fermi surface in the ferromagnet can have a role in modifying the character of the interface states, we consider in the present analysis the hopping terms $t_x=t_y=t=1.0$ to be uniform along the $x$- and $y$- directions in the FM side.\\

For the analysis of superconducting state in the heterostructure, the Hamiltonian in Eq.~\ref{eq.hamiltonian} is 
decoupled within the Hartree-Fock approximation as 
\begin{eqnarray}
V^{\sigma \sigma{^{'}}} n_{\mathbf{i}\sigma} n_{\mathbf{j}\sigma{^{'}}} \simeq && V^{\sigma \sigma{^{'}}} (\Delta^{\sigma\sigma{^{'}}}_{\mathbf{i}\mathbf{j}} c^{\dagger}_{\mathbf{j} \sigma} c^{\dagger}_{\mathbf{i} \sigma{^{'}}} \\
&+&\bar{\Delta}^{\sigma \sigma{^{'}}}_{\mathbf{i}\mathbf{j}}c_{\mathbf{i}\,\sigma{^{'}}} c_{\mathbf{j}\,\sigma}-|\Delta^{\sigma \sigma{^{'}}}_{\mathbf{i}\mathbf{j}}|^2), \nonumber
\end{eqnarray}
\noindent where the general pairing amplitude on a bond between spin $\sigma$ and $\sigma'$ electrons at the 
sites ${\mathbf{i}}$ and ${\mathbf{j}}$ is given by $\Delta_{\mathbf{i}\mathbf{j}}^{\sigma \sigma{^{'}}}=\langle c_{\mathbf{i}\,\sigma} c_{\mathbf{j}\,\sigma{^{'}}} \rangle$. Spin-triplet order parameters can be expressed in a matrix form as
\begin{eqnarray*}
\Delta(k)=
\left(\begin{array}{cc}
  \Delta_{\uparrow\uparrow}(k) & \Delta_{\uparrow\downarrow}(k)\\
  \Delta_{\downarrow\uparrow}(k) & \Delta_{\downarrow\downarrow}(k)
\end{array}\right)
= \left(\begin{array}{cc}
  -d_x+i d_y & d_z \\
  d_z & d_x+id_y
\end{array}\right) \, ,
\end{eqnarray*}
\noindent where the $\vec{d}$-vector components are related to the
pair correlations for the various spin-triplet configurations having zero spin projection 
along the corresponding symmetry axis. The three components
$d_x=\frac{1}{2}(-\Delta_{\uparrow\uparrow}(k)+\Delta_{\downarrow\downarrow}(k))$,
$d_y=\frac{1}{2 i}(\Delta_{\uparrow\uparrow}(k)+\Delta_{\downarrow\downarrow}(k))$
and $d_z=\Delta_{\uparrow\downarrow}(k)$ are expressed in terms of
the equal spin $\Delta_{\uparrow\uparrow}(k) ~\mathrm{and}~
\Delta_{\downarrow\downarrow}(k)$, and the anti-aligned spin
$\Delta_{\uparrow\downarrow}(k)$ pair potentials. For the cases
upon examination, the pairing interaction $V$ is assumed to be non zero in the
$\uparrow\uparrow$ and $\downarrow\downarrow$ channels and, thus, $\Delta_{\sigma\sigma}(k)$ are the only non-vanishing order
parameters. 
This implies that the $\vec{d}$-vector lies in the $xy$-plane, which is
chosen to be coincident with the $xy$-plane of the heterostructure as indicated in Fig.~\ref{fig.100.fm.ttsc.interface}.

\begin{figure}[!t]
\begin{center}
\includegraphics[width=0.45\textwidth]{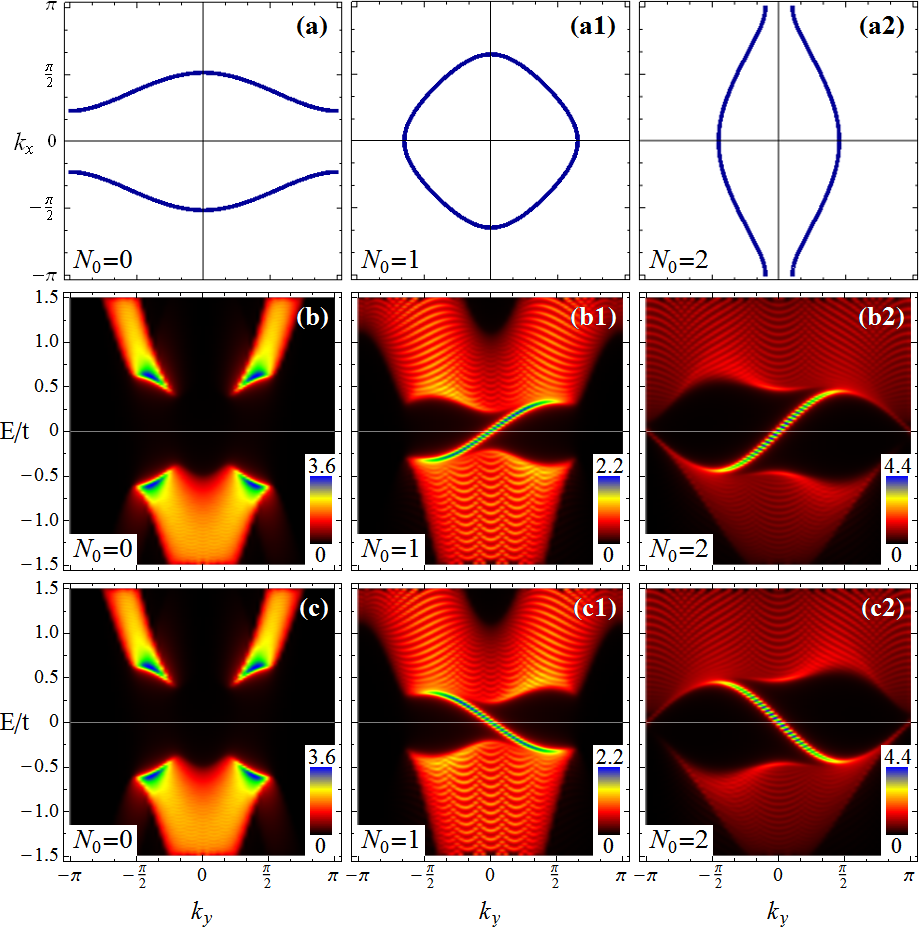}
\end{center}
\caption{(color online). Top panels: Fermi surface of the normal state of the spin-triplet superconductor associated with
different number $N_0$) of Majorana helical modes at its edge. 
The parameters are $t_x=0.4\,t$, $t_y=1.0\,t$ and 
$\mu=-0.7\,t$ for $N_0=0$, $t_x=1.0\,t$, $t_y=1.0\,t$ and $\mu=-1.1\,t$ for $N_0=1\,t$, and $t_x=0.4\,t$, $t_y=1.0\,t$ and $\mu=-1.1\,t$ for $N_0=2$.
Contour map of the spin-up (middle panels) and spin-down (bottom panels) spectral function at the edge of the spin-triplet superconductor 
for the corresponding Fermi surfaces.}
\label{fig.fermi.surface.and.spectrum}
\end{figure}

To design the 2D spin-triplet superconductors with time-reversal invariance, 
we exploit the fact that the number of Majorana helical modes is
intimately linked to the topological invariants of the Fermi
surface in the normal state.\cite{Sato2009} Hence, by suitably choosing the 
topology of the Fermi surface for the considered single-band model, it is possible to obtain a superconducting state with $N_0=0,1,2$ edge states.
These are obtained for three different topology of the Fermi surface as depicted in the top panels of Fig.~\ref{fig.fermi.surface.and.spectrum} and for which 
the hopping parameters for the $x$ and $y$ directions 
are (a) $t_x=0.4\,t$, $t_y=1.0\,t$, $\mu=-0.7\,t$ ($N_0=0$), 
(b) $t_x=1.0\,t$, $t_y=1.0\,t$, $\mu=-1.1\,t$ ($N_0=1$) and 
(c) $t_x=0.4\,t$, $t_y=1.0\,t$, $\mu=-1.1\,t$ ($N_0=2$), respectively.

The numerical analysis consists in evaluating self-consistently the pair correlations
\begin{eqnarray}
\Delta^{\sigma \sigma}_{p_{x(y)}}(\mathbf{i})=\Delta^{\sigma\sigma}_{\mathbf{i},\mathbf{i}+\text{\^{x}}(\text{\^{y}})} \label{eq.delta.px}  
\end{eqnarray}
by solving the Bogoliubov-De Gennes equations using the Hamiltoninan defined in Eq.~\ref{eq.hamiltonian}. By choosing the electronic parameters mentioned above, the superconducting region exhibit a stable spin-triplet state with helical $\vec{d}=(p_y,p_x,0)$ symmetry. All the results discussed in this paper are obtained at zero temperature, for a lattice size $L_x=L_y=80$ sites and by choosing a pairing interaction $V=-2.0t$ such that the bulk gap $V \cdot \Delta$ is still small compare to the spectrum bandwidth. As we shall see, the results depend essentially on the modification of the helical edge states at the interface and, thus, do not depend substantially on $V$.\\

Before considering the proximity with the ferromagnet, it is instructive to investigate the edge states of the spin-triplet superconductor interfaced with the vacuum by solving the Hamiltonian in Eq. (1) without the ferromagnet and for open boundary conditions along the $x$-direction. The site- and spin-dependent spectral function at any given distance from the interface with the vacuum are obtained from the self-consistent evaluation of the TTSC order parameter by evaluating the imaginary part of the single particle Green's function through the Fourier transformation of the two-time correlator

\begin{eqnarray}
A_{\sigma}(i_x,k_y)(\omega)=-\frac{1}{\pi} \int d\tau {\rm Im}[\langle c^{\dagger}_{i_x k_y \sigma}(\tau) c_{i_x k_y \sigma}(0)\rangle] e^{i\,\omega \tau} \, \nonumber
\\
\end{eqnarray}

\noindent for which we have assumed periodic boundary conditions along the $y$-direction and where $\langle ...\rangle$ is the average on the ground state and $c^{\dagger}_{i_x,k_y \sigma}$ the creation operator of an electron at the site $i_x$ with momentum $k_y$ and spin-polarization $\sigma$. Hereafter, for convenience the $\uparrow (\downarrow)$ configurations will refer to the
$z$-direction in the spin space are then perpendicular to the $\vec{d}$-vector.

In Fig.~\ref{fig.fermi.surface.and.spectrum} we report the density plot of the spin-up (middle panels) and spin-down (bottom panels) spectral functions evaluated at the boundary of the TTSC, i.e. at the site $i_x=1$, where the weight of the helical modes in the energy spectrum is maximal. Nevertheless, the edge states are located inside a region of about one superconducting coherence length $\xi_{sc}$ close to the interface, which is of the order of 20 atomic distances for the given value of the superconducting pairing coupling $V$.
Let us first analyze the spectral functions in the configuration with $N_0=1$ gapless mode obtained for isotropic hopping amplitudes within the lattice. For both, the spin-up and spin-down channels, we can clearly identify the superconducting gap in the spectrum in the energy range $-0.25\,t<E<\,0.25t$ as well as midgap states. The latter are gapless at $k_y=0$ and almost linearly dispersing. Furthermore, the helical modes connect the edges of the continuum band above and below the Fermi level at about $k_y=-\pi/2$ and $k_y=\pi/2$. We note that the dominant spectral weight resides in the helical modes which are fully spin polarized and that, due to the time reversal constraint, are symmetrically linked by inverting the direction of the momentum, i.e. $A_{\ua}(i_x,k_y)(\omega)=A_{\da}(i_x,-k_y)(\omega)$.
When considering an open Fermi surface along the $k_y$ direction, see (a2), the midgap helical modes have a different structure and exhibit two gapless branches crossing at the momenta $k_y=0$ and $\pi$ of the Brillouin zone. The spectral weight distribution is however not symmetric with a larger amplitude close to $k_y=0$.
The case of a Fermi surface that is open along the $x$-direction, see (a), yields midgap states which are fully gapped and confined close to the gap edge of the spectrum in the momentum window around $k_y = \pm \pi/2$.

The Fig.~\ref{fig.fermi.surface.and.spectrum} shows that for a given spin
polarization the dispersion $\omega(k_y)$ of the midgap states is odd in momentum. Therefore, 
the spin-polarized configurations can sustain a flowing current along the boundaries of the TTSC. This is an important and the main physical quantity for the present study which we aim to investigate in the proximity of the ferromagnetic interface. The $\sigma$ spin-polarised kinetic currents flowing along the TTSC interface at the site $i_x$, having a spin polarization along the $\alpha$ direction ($\alpha=x,y,z$), is expressed as

\begin{equation}
J^{\alpha}_{\sigma}(i_x)= \frac{2\,t}{L_y} \sum_{k_y} \sin(k_y) \langle c^{\dagger}_{i_x, k_y \sigma} c_{i_x k_y \sigma} \rangle.
\label{eq.J}
\end{equation}

\noindent The charge current is hence obtained by summing the contribution of the $\uparrow$ and $\downarrow$ spin-polarized electrons as $J_{c} (i_x)=J^{z}_{\uparrow}(i_x)+J^{z}_{\downarrow}(i_x)$ while the $z$-polarised spin current is obtained from their subtraction as $J^{z}_{s} (i_x)=J^{z}_{\uparrow}(i_x) - J^{z}_{\downarrow}(i_x)$.  As we have seen in Fig.~\ref{fig.fermi.surface.and.spectrum}, the time reversal symmetry insures the dispersion of the spin-up and spin-down gapless modes, for both \None and \Ntwo cases, to be opposite. Hence, spin-up and spin-down electron are counter-propagating, i.e. $J^{z}_{\ua}(i_x)=-J^{z}_{\da}(i_x)$, which leads to the existence of a finite $z$-polarised spin current along the TTSC interface while the charge current is perfectly cancelled, i.e. $J_{c}(i_x)=0$.


\section{Evolution of the energy spectrum at the FM-TTSC interface}

In this section we discuss the evolution of the electronic states at the interface between the 
FM and the TTSC, both inside and above the superconducting energy gap, for the cases with one and two pairs of Majorana helical modes.
Such analysis is particularly relevant to understand how the degree of mixing between the
midgap edge modes and the magnetic states of the FM close to the Fermi level is interrelated 
to the variation of the spin and charge currents. We shall consider an orientation 
of the magnetic exchange both perpendicular and coplanar to the $\vec{d}$-vector. We expect the effects on the
helical modes to be significantly distinct because a magnetization parallel to the $\vec{d}$-vector is generally pair breaking for the spin-triplet superconducting state.

\subsection{Magnetization perpendicular to the $\vec{d}$-vector}

\begin{figure}[!t]
\begin{center}
\includegraphics[width=0.47\textwidth]{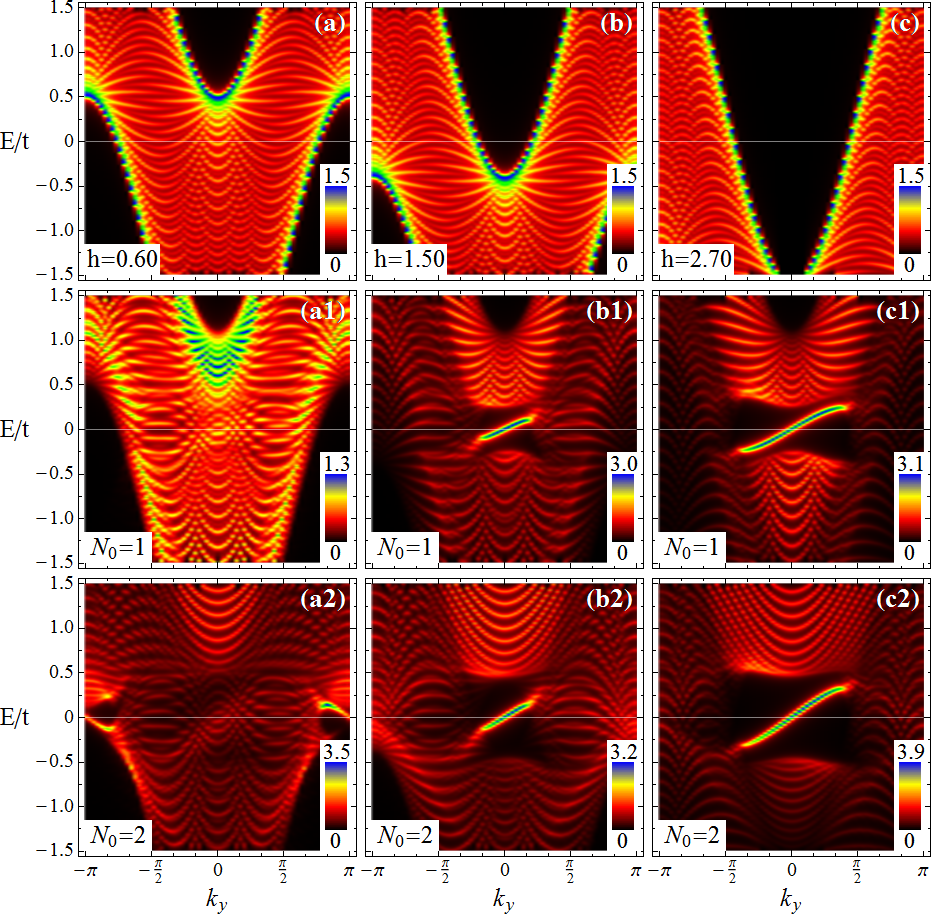}
\end{center}
\caption{(color online).Energy spectrum for spin-up electrons at three representative values of the exchange field $h=0.6 t$ (left), $1.50 t$ (middle) and $2.70 t$ (right) assuming that the direction is perpendicular to the $\vec{d}$-vector ($\theta=0$ and $\phi=\pi/2$). They are determined at a position $i_x$ that is in the bulk of the FM (top) and at the edge of the TTSC for $N_0=1$ (middle) and $N_0=2$ (bottom).}
\label{fig.spectrum.spin.up}
\end{figure}

\begin{figure}[!t]
\begin{center}
\includegraphics[width=0.47\textwidth]{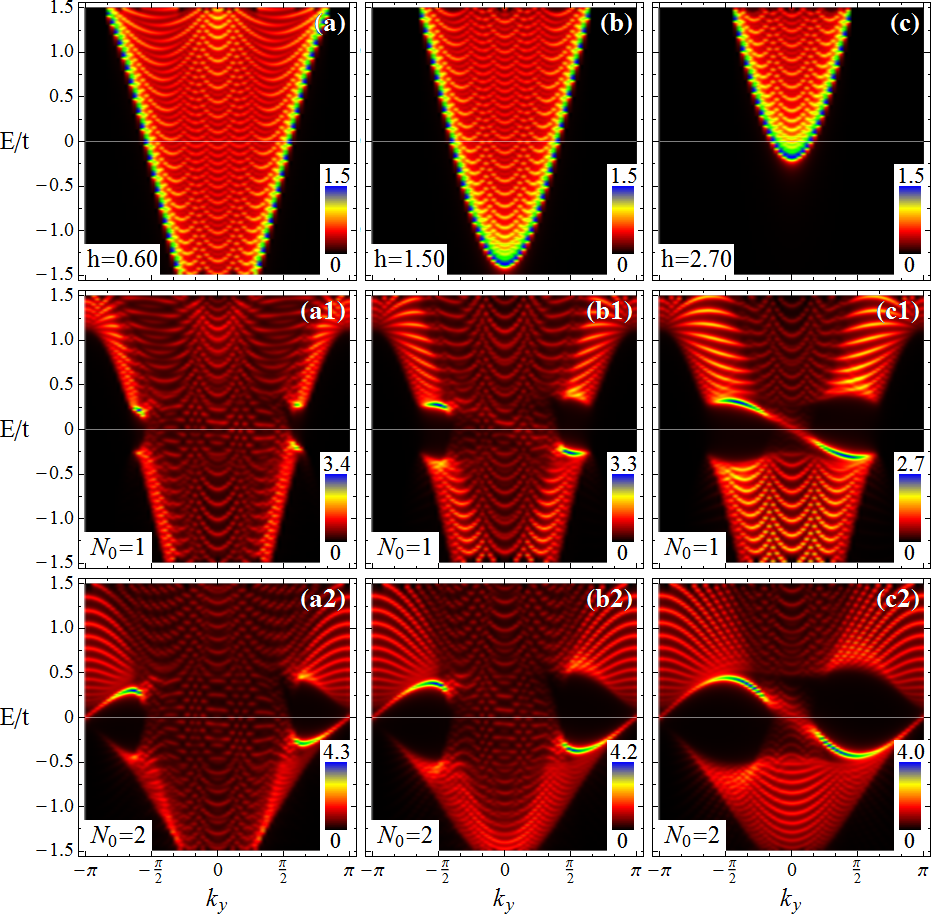}
\end{center}
\caption{(color online).Energy spectrum for spin-down polarization at three representative values of the 
exchange field, i.e. $h=0.6t$ (left), $1.50 t$ (middle) and $2.70 t$ (right), 
assuming that the direction is perpendicular to the $\vec{d}$-vector ($\theta=0$ and $\phi=\pi/2$). They have been obtained at a position in the interior of the FM (top) and at the edge of the TTSC for $N_0=1$ (middle) and $N_0=2$ (bottom).}
\label{fig.spectrum.spin.down}
\end{figure}

The first analysis concerns the case with an exchange field parallel to the
$z$-direction, thus perpendicular to the $\vec{d}$-vector. We report in Figs.~\ref{fig.spectrum.spin.up} and~\ref{fig.spectrum.spin.down} the spectral functions for the spin majority (spin-up) and minority (spin-down) components for three values of the ferromagnetic exchange $h$. They are chosen to be representative of the regimes of weak ($h=0.6\,t$), intermediate ($h=1.5\,t$) and strong ferromagnet ($h=2.7\,t$), respectively, in order to single out the role of the magnetization's strength in controlling the modification of the helical modes in the TTSC. The upper row panels in Fig.~\ref{fig.spectrum.spin.up} and~\ref{fig.spectrum.spin.down} depict the ferromagnetic single particle spectral functions obtained inside the ferromagnetic bulk while the middle and lower rows show the spectral functions obtained at the FM-TSC interface, i.e. at the site $i_x=1$, for the \None and \Ntwo superconductors, respectively.\\

One can note that, as expected from the variation of the occupation number for each spin in the Brillouin zone along the direction of the exchange field, the distribution of the electronic states around the Fermi level, as a function of the transverse momentum $k_y$,
manifests an energy splitting between the spin-up and spin-down channels.
The features of the electronic structure in the FM are tied to the considered 2-dimensional tight-binding model, 
for which the spectrum is $\varepsilon (k_x,k_y)=-2 t[\cos(k_x) + 2 t\cos(k_y)] -\mu$, thus having a topology of the Fermi surface that is closed (electron-like) around the center of the Brillouin zone.
Then, projected on the conserved momentum along the $y$-direction, the increase of the spin density leads to a reduction of the electronic states around the point at $k_y=0$ at low energy and an increase of the
electronic states close to the zone boundary ($k_y=\pi$) both at low and high energies exhibiting a continuum of excitations
that extends up to about half of the bandwidth.
Such a variation of the electronic distribution reflects the change of the electronic structure for the majority and the
minority spin electrons. For a given total electron density the effective Fermi level $k^{F}_{y\uparrow}$ depends on $h$ for
each projected spectrum and separates two different regions of the Brillouin
zone. In the range $[-k^{F}_{y\uparrow},k^{F}_{y\uparrow}]$
there are no states available for the spin majority electrons close to the zero energy because the bottom of the band is at $k_y=0$
and it gets lowered in energy by the exchange field (see Fig.~\ref{fig.spectrum.spin.up}) to allow the increment of the spin-majority electron density.
Outside this range of moments, the electronic states are accessible in a continuum of energies with a bandwidth that grows with the amplitude of $h$.
The evolution is opposite for the spin minority electrons. As one can see in top panels of Fig. ~\ref{fig.spectrum.spin.down} already at $h=0.6 t$
the effective $k^{F}_{y\downarrow}$ is close to $k_y=0$ with a distribution of electronic states close to the Fermi level that tends to be 
vanishing at all $k$ points in the Brillouin zone. The further increase of $h$ in the FM leads to a shrinking of the
window in the Brillouin zone where there are allowed occupied states and they become more and more concentrated uniquely around the $k_y=0$ point.
The energy bandwidth is also reduced as one is approaching the half-metallic regime where the spin-minority electron density
tends to zero.
Hence, the spectral-function evolution clearly shows that the electronic states close to zero energy accumulate
around the zone boundary (center) of the Brillouin zone for the spin majority (minority) electrons as the exchange field
is varied from zero to the half-metallic amplitude. 
Moreover, the bandwidth of the low energy states grows (decreases) as a function
of the exchange field for the majority (minority) spin electrons. Such interrelation holds for 
an electron-like Fermi surface and it is reversed if one considers a different dispersion in the FM with a hole-like type of Fermi surface.\\

Taking into account these features, one can closely analyze how the edge states in the
TTSC region get modified by the presence of the FM. 
Let us start by the $N_0=1$ case for which two spin-polarized helical modes with opposite velocity crossing zero energy at $k_y=0$ are present inside the gap. Then, according to the distribution of the electronic spectrum in the FM, we expect the spin-minority channel to be
more affected than the spin-majority one, at least at large values of the exchange field.
This result is confirmed by inspection of the Figs.~\ref{fig.spectrum.spin.up} and~\ref{fig.spectrum.spin.down} showing the
dependence of the spectral functions in the superconducting region of the heterostructure.

For $h=0.6 t$, we can observe for both spin-up and spin-down spectral functions that the gap is suppressed and that the hybridization with the magnetic states in the FM destroys the midgap states, see (a1). 

The evolution of the electronic structure inside the superconducting gap follows
directly that of the spectra in the ferromagnetic region. For the majority spin electrons,
the edge modes around the centre $k_y=0$ of the Brillouin zone gets more robust
as the magnetization grows up to the half-metallic limit, Fig.~\ref{fig.spectrum.spin.up}(b1) and~(c1). Moreover, the large distribution
of spectral weight at the zone boundary persists in such a way that the gap is suppressed there and the spectral weight of the edge states is also renormalized down to zero.
On the other hand, for the minority spin electrons there is a substantial renormalization
of the spectral weight for the edge modes even when the superconducting gap is clearly visible. Due to the distribution of the electronic structure in the FM it is the part of the edge states which joins the continuum above the superconducting gap that acquires spectral weight in the strong ferromagnet regime while the spectrum close to $k_y\sim 0$ becomes more coherent though with a reduced occupation probability, Fig.~\ref{fig.spectrum.spin.down}(b1) and~(c1).
When comparing the edge modes in Figs.~\ref{fig.spectrum.spin.up} and ~\ref{fig.spectrum.spin.down} with that of the interface to the vacuum, one can observe that the coupling with the FM results into a significant renormalization of the helical-modes spectral weight, with a minor change of the dispersion mainly related to the increase of the effective velocity at large momenta close the gap edge.

Concerning the case \Ntwo, the hybridization of the electronic states with momentum $-\pi/2<k_y<\pi/2$ is similar to the case \None. 
However, differences arise at momentum $k_y=\pm \pi$ since, in this case, additional gapless edge states are present. 
For spin up electrons, we can observe in Fig.~\ref{fig.spectrum.spin.up}(a2) that the states are unmodified around $k_y=\pm \pi$ at small values of $h$. 
Then, they hybridize completely with the ferromagnet states for intermediate and large values of the exchange field, Fig.~\ref{fig.spectrum.spin.up}(b2) and~(c2).
Finally, as a distinct feature of the helical modes variation at the interface, the spectrum for the minority spin electrons in Fig.~\ref{fig.spectrum.spin.down}(a2-c2)is completely unaffected around $k_y=\pm \pi$ at any value of the exchange field.

\subsection{Magnetization coplanar to the $\vec{d}$-vector}

The behavior of the spectral function is more complex when the exchange field is coplanar to the $\vec{d}$-vector because such 
an orientation can be pair breaking and transverse to the spin configuration 
of the superconducting state. 
Furthermore, due to the structure of the $\vec{d}$-vector with orbital inequivalent 
$x$ and $y$ components, i.e. $\vec{d}=(p_y,p_x,0)$, 
the orientation of the magnetic exchange can lead to significant in-plane anisotropy 
in the edge electronic spectrum.
Since for a magnetic exchange that lies in the $xy$-plane of the spin-space the up and down orientations
are equivalent, i.e. $A_{\ua}(i_x,k_y)(w)=A_{\da}(i_x,k_y)(w)$ for $i_x \in \text{FM}$, and due to the existence of the time reversal symmetry which insures $A_{\ua}(i_x,k_y)(w)=A_{\da}(i_x,-k_y)(w)$ for $i_x \in \text{TTSC}$, we limit the study of the energy spectrum to one spin polarization. Moreover, we discuss the spectral function corresponding to \Ntwo helical modes as it presents all the relevant features arising from a coplanar magnetic exchange field, the case with \None modes can be deduced by direct inspection of the structures close to $k_y\sim 0$.

\begin{figure}[!t]
\begin{center}
\includegraphics[width=0.47\textwidth]{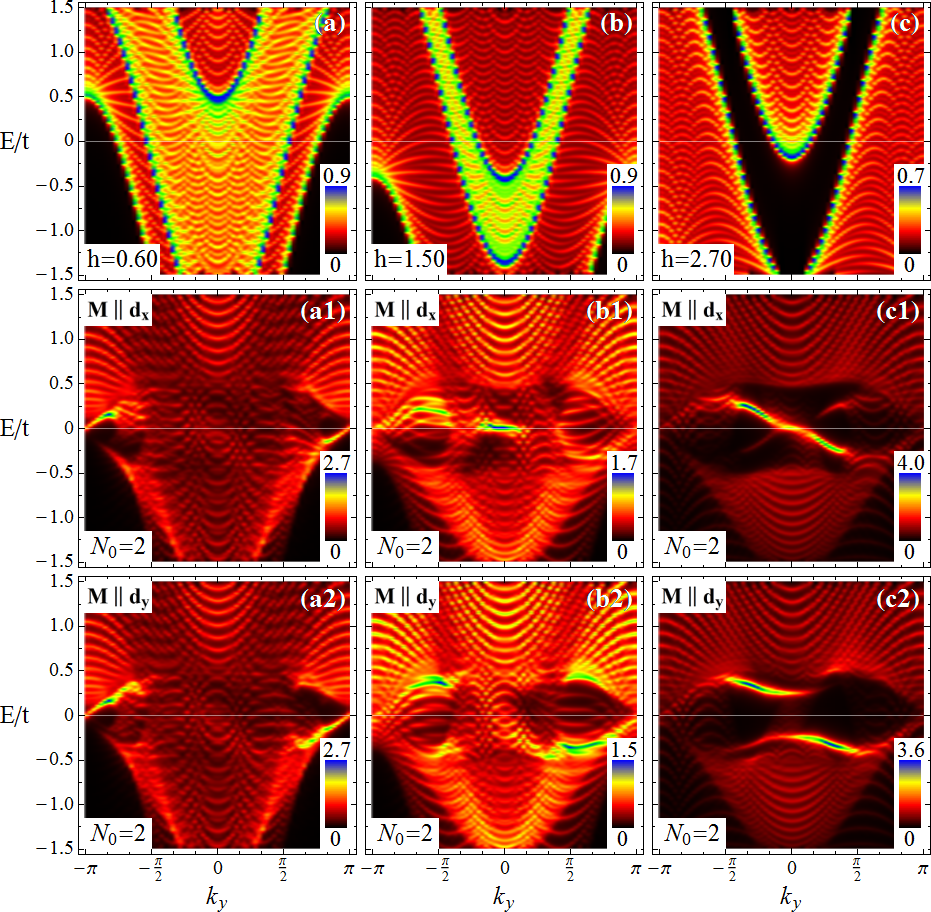}
\end{center}
\caption{(color online).Energy spectrum for a given spin-polarization coplanar to the 
$\vec{d}$-vector at $h=0.6 t$ (left), $1.50 t$ (middle) and $2.70 t$ (right) assuming that the exchange field is coplanar to the $\vec{d}$-vector ($\theta=\pi/2$ with $\phi=0$ and $\pi/2$). They are obtained in the FM bulk (top row panels) and at the nearest site $i_x=1$ to the FM-TTSC interface (bottom row panels) for the configuration $N_0=2$.}
\label{fig.spectrum.spin.down.t.90}
\end{figure}

In Fig. \ref{fig.spectrum.spin.down.t.90} we compare the spin-down spectral functions for two magnetic exchange orientations along the $x$- and $y$-direction moving from weak to half-metallic ferromagnet.
The electronic spectra in the ferromagnet (top row) presents a two-band structure that reflects the 
nonzero spectral weight of both the up and down split bands when the exchange is coplanar to the 
$\vec{d}$-vector. This implies that, already at small values of the exchange $h$, there is a distribution of spectral weight close to the Fermi level at any value of $k_y$ which makes possible a mixing with all the midgap helical modes.
For the weak ferromagnet regime, i.e. $h=0.6 t$, the degree of magnetic quasiparticle poisoning makes the gap undetectable and one cannot identify significant differences when switching the exchange from $x$- to $y$-direction, see Fig~\ref{fig.spectrum.spin.down.t.90}(a1) and~(a2). In this regime the helical modes close to $k_y \sim \pi$ do not hybridize much with the magnetic states and a slight $xy$ asymmetry in the spectral weight can be observed with more coherent intensity along the $y$ direction associated with the $\text{sin}(k_x)$ orbital pairing symmetry.

More evident anisotropies emerge by analyzing the behavior at larger values of the exchange 
($h=2.7t$) when, as a consequence of an inverse proximity effect, the edge states acquire a spin polarization that 
is more robust, Fig~\ref{fig.spectrum.spin.down.t.90}(a1) and~(a2). Then, due to the up-down mixing induced by leaking of the magnetization in the superconductor, the spectral function for a given spin polarization exhibits shadows of the helical modes with opposite spin configuration as well. 
The role of mixing is considerably relevant when comparing the case of a ferromagnet with 
$x$ or $y$ orientations.
For the exchange parallel to the $x$ directions, the up-down mixing strengthens the intensity of the helical modes close to $k_y=0$ leading to a flattening of the dispersion (Fig. \ref{fig.spectrum.spin.down.t.90}(c1)).
On the other hand, a spin-polarization parallel to the $y$-component of the $\vec{d}$-vector
couples to the orbital component of the superconducting state that leads to the zero energy states. This implies a splitting helical modes with a gap opening at $k_y=0$ and a dramatic modification of the dispersion (Fig. \ref{fig.spectrum.spin.down.t.90}(c2)).


\section{Currents at the interface: role of magnetization orientation and number of helical modes}

In this section we discuss, for both cases of one- and two-pairs of Majorana helical modes, the dependence of the spin- and charge-currents at the interface with the ferromagnet by varying the strength and the orientation of the 
exchange field. The analysis of the currents will be closely traced by considering the changes of the electronic 
spectra at a given spin-polarization as presented in the previous Section. 
The main question we address is how the number of helical modes influences the 
spatial dependence and the currents flowing at the interface. 
The mixing between the edge states and the ferromagnetic states close to the interface is the key dynamical parameter 
that controls the currents behavior at the interface. As a general feature, due to the breaking 
of time reversal symmetry in the ferromagnet, the magnetic-helical hybridization leads to an unbalance between 
the spin-resolved currents $J_{\ua}$ and $J_{\da}$ and thus to nonvanishing charge currents flowing at the interface.
Moreover, due to the symmetry of the superconducting state the dominant currents are spin-polarized along the direction 
perpendicular to the $\vec{d}$-vector.

\subsection{Magnetization perpendicular to the $\vec{d}$-vector}

\begin{figure}[!t]
\begin{center}
\includegraphics[width=0.43\textwidth]{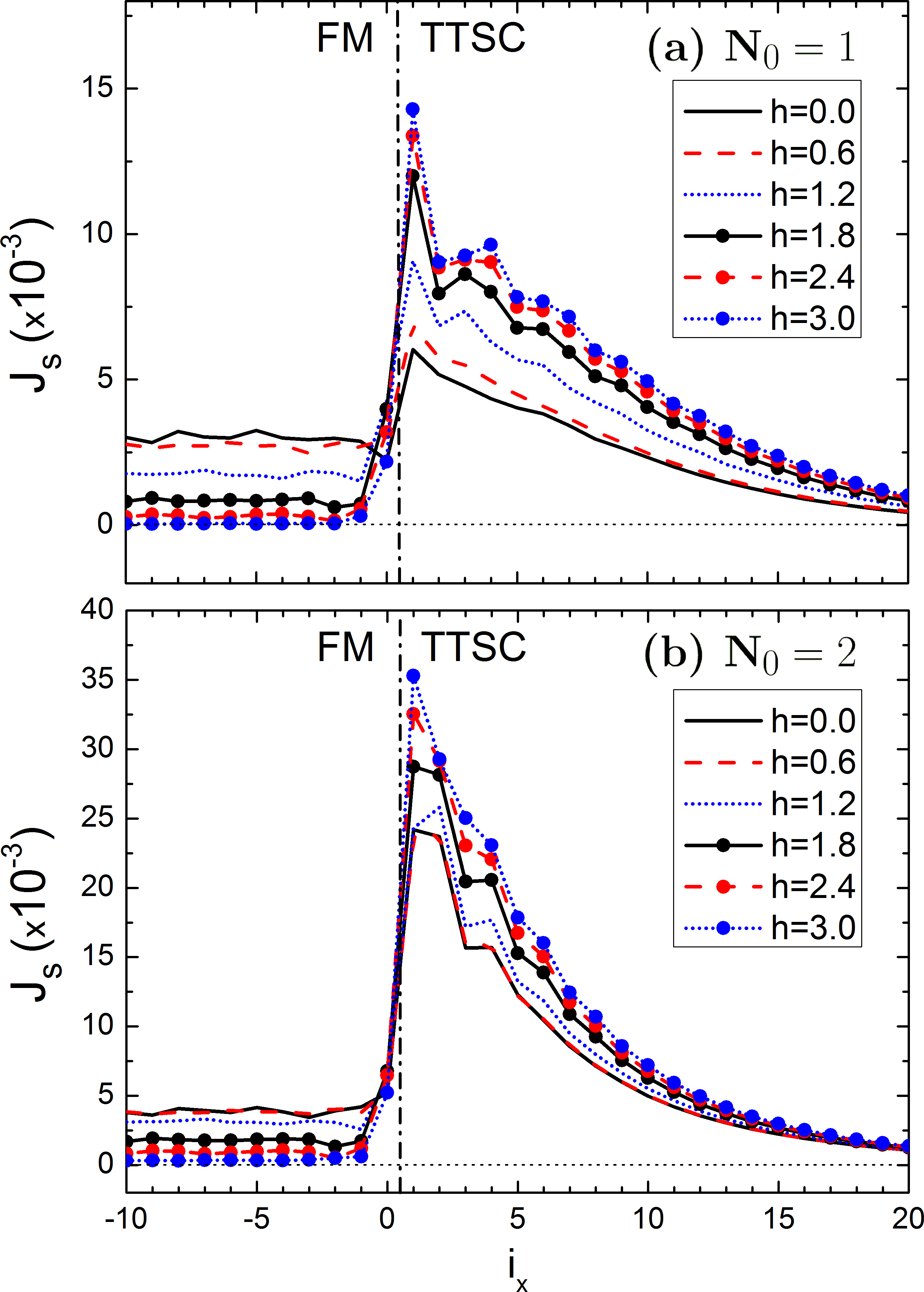}
\end{center}
\caption{Spatial variation of the spin current $J_s(i)$ along the FM-TTSC interface for various strengths 
of the exchange field $h$ for (a) one and (b) two helical modes. The magnetization is 
perpendicular to the ${\vec d}$-vector, i.e. $\theta=0$. The dashed-dotted line indicates the interface between the FM and TTSC regions.}
\label{fig.Jspin.t0}
\end{figure}

We start the discussion by considering the case of a magnetization in the FM that is perpendicular to the $\vec{d}$-vector. 
For such configuration, the spatial variation of the \None and \Ntwo spin current flowing along the FM-TTSC interface is presented in Fig.~\ref{fig.Jspin.t0} 
at different values of the exchange field, moving from the unpolarized normal state, i.e. $h=0$, until the half-metallic regime at $h=3.0\,t$. 

As one can note, the general trend is similar for both cases, \None (Fig.~\ref{fig.Jspin.t0}(a)) and \Ntwo (Fig.~\ref{fig.Jspin.t0}(b)). 
Indeed, we observe that the spin current is maximal at the interface 
on the side of the TTSC, and then decays moving into the FM and further into the superconducting region. 
Within the superconductor, 
the characteristic length for which the spin current is suppressed is set by the coherence length of the superconductor 
that for the chosen value of the pairing amplitude is $\xi_{sc}\sim 20$ in unit of the atomic sites distance. 
This spatial domain in the TTSC corresponds to that where the edge states in the spectrum have a 
nonvanishing spectral weight. 
At larger distances from the interface, i.e. $i_x>\xi_{sc}$, the spectrum turns out to be fully gapped and the 
spin currents is suppressed to zero.

Nevertheless, by close inspection of the profile of the spin-current, we can observe distinct features between the two cases. We find that the spin-current grows faster for the case \None than for \Ntwo when moving from the bulk TTSC to the interface because the presence of additional helical modes in the latter allows for a larger penetration in the inner side of the superconductor.
We point out that the characteristic length of the spatial variation of the spin current does not seem to be
significantly modified if compared to the case obtained at the vacuum-TTSC interface.
On the other hand, the penetration of the spin currents in the FM region goes through 
an abrupt jump at the interface and it can 
stay into the ferromagnet almost unchanged on the scale of the proximity coherence length. 
It is worth mentioning that, for a magnetization perpendicular to the $\vec{d}$-vector, the exchange is not pair breaking and the 
proximity scale is basically set in the clean limit by the proximity coherence length as given by the
ratio between the Fermi velocity and the temperature, i.e. $\sim v_F/T$. 
Hence, since we are dealing with a ballistic ferromagnet, 
the spin current can be sustained on long distances from the interface at low temperature. 
Approaching the half-metallic regime, due to the large suppression of the Andreev processes
close to the interface, the spin current gets strongly reduced and confined close by the interface. \\

In order to further analyze the evolution of the spin current in the TTSC region and to extract possible differences 
in the capacity of carrying currents in terms of the number of helical modes, we compute the integrated current densities for spin-up and spin-down electrons evaluated within the TTSC region. Following Eq.~\ref{eq.J}, the expressions for the total spin and charge currents are

\begin{eqnarray}
J^{tot}_{\sigma} &&= \sum^{}_{i_x \in TTSC } J_{\sigma}(i_x), \\
J^{tot}_s &&= J^{tot}_{\ua} - J^{tot}_{\da},\\
J^{tot}_c &&= J^{tot}_{\ua} + J^{tot}_{\da} .
\end{eqnarray}
%

\noindent Hence, $J^{tot}_s$ and $J^{tot}_c$ represent the total spin and charge currents in the TTSC, respectively, flowing along the interface, i.e. in the $y$-direction. All the results concerning the total currents shown in the following, for both cases \None and \Ntwo, are scaled to the value of the total spin current $J^{0}_{s}$ at the \None and \Ntwo superconductors, respectively, interfaced with the vacuum. They verify the relation $J^{0}_{s}(~\text{\Ntwo}) \simeq 2 \cdot J^{0}_{s}(~\text{\None})$.

\begin{figure}[!t]
\begin{center}
\includegraphics[width=0.40\textwidth]{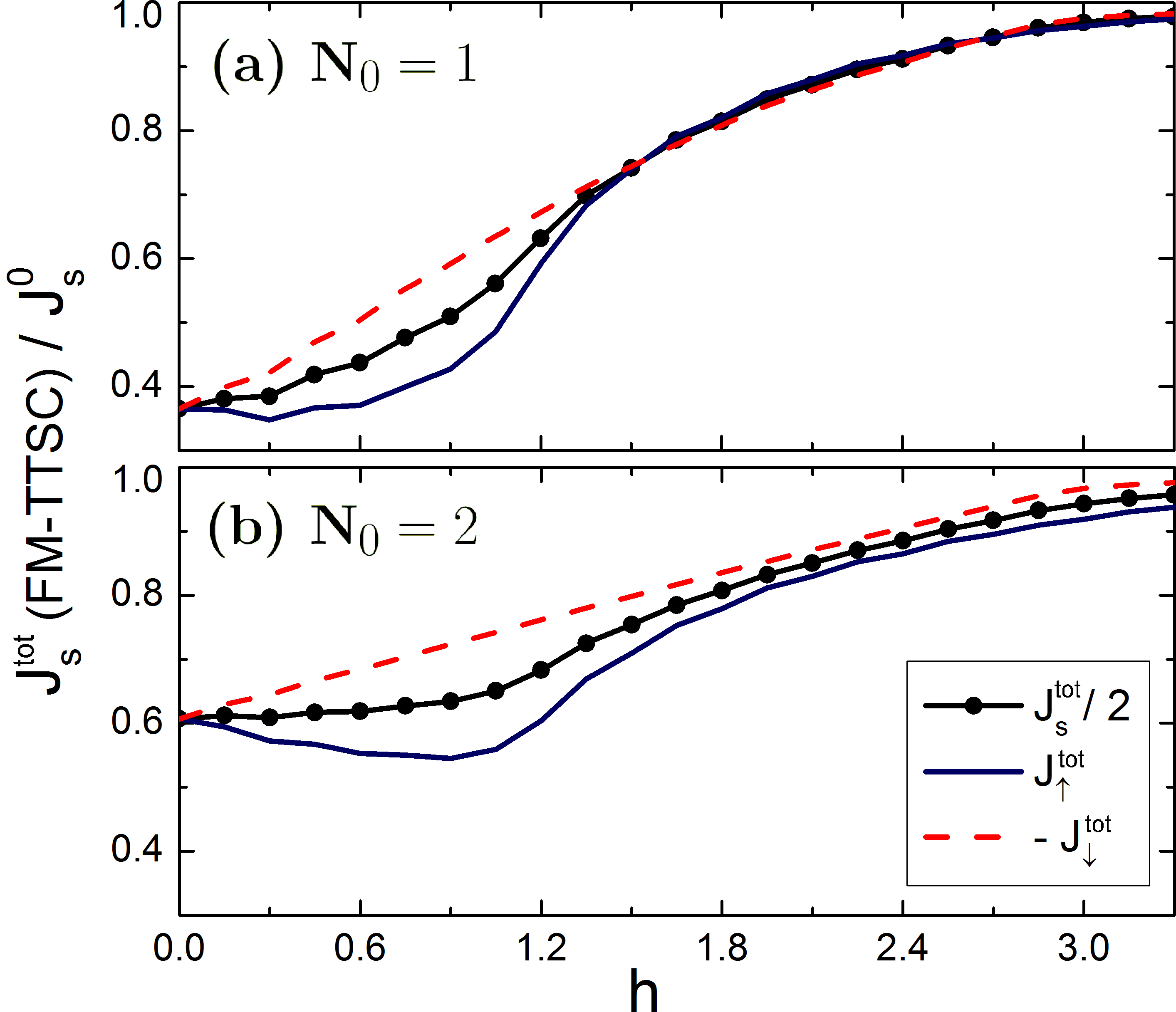}
\end{center}
\caption{Integrated spin current $J_s(i)$ within the TTSC region for various strengths of the exchange field $h$ and 
normalized to that obtained at the vacuum-TTSC interface. The results are showed for configurations with $N_0=1$ (a) and $N_0=2$ (b) helical modes.
The magnetization is perpendicular to the $\vec{d}$-vector, i.e. $\theta=0$. }
\label{fig.Jspin.t0.sum}
\end{figure}

The variations of $J^{tot}_{\ua}$(solid blue), -$J^{tot}_{\da}$(dashed red) and $J^{tot}_s$(dotted black) as a function of the exchange $h$ are shown in Fig. \ref{fig.Jspin.t0.sum}. As we can see, the presence of the FM breaks the up-down symmetry in the current, and
the majority spin electrons in general exhibit a smaller capacity to carry current when a magnetization 
is put in proximity of the helical superconductor.
This asymmetry tends to reduce down to zero as the ferromagnetism 
becomes stronger although the asymptotic behavior turns out to be different for the 
case of \None and \Ntwo helical modes. Such behavior 
can be understood by taking into account how the mixing of the
helical and magnetic states occur in the Brillouin zone according to the 
analysis of the spectral function presented in Sec. III. 
For \None at larger values of the exchange field, i.e. above $h\sim 1.5\,t$, the low energy helical modes around $k_y\sim 0$, for 
both spin polarizations, are not any longer hybridized 
with the magnetic states (see Figs.~\ref{fig.spectrum.spin.up} and~\ref{fig.spectrum.spin.down}) and thus the up-down symmetry is about to be 
recovered, with a small residual difference mainly arising from the contributions at energies above the superconducting gap. 
On the other hand, 
the \Ntwo helical superconductor has spin-current contributing modes at $k_y\sim 0$ and at momenta close to 
the zone boundary $\pm \pi$. 
The increase of the exchange field can avoid the magnetic-midgap states mixing 
nearby $k_y\sim 0$ in both spin channels, nevertheless
it is not possible to get rid of the hybridization 
at the zone boundary between the helical modes and the
majority spin channel. This residual single-particle poisoning of the helical modes will 
keep a nonvanishing asymmetry between the majority and minority spin currents 
even in the half-metallic ferromagnet regime.

Let us discuss in more details on the evolution of the integrated spin-current for the case of \None helical modes. 
The evolution of the spin current identifies two different regimes in terms of the strength of the ferromagnet. 
Starting from the unpolarized configuration, i.e. $h=0$, we find that the spin current is more than half suppressed if compared with the amplitude at the vacuum-TTSC interface.
Then, in the range from weak to intermediate ferromagnet, i.e. moving from $h=0$ to about $h=1.0 t$, 
we can observe that $J^{tot}_{\da}$ grows linearly while $J^{tot}_{\ua}$ remains almost constant. 
These variations can be understood by inspection of the spectral functions in Figs.~\ref{fig.spectrum.spin.up} and~\ref{fig.spectrum.spin.down}. 
Indeed, for small values of $h$, we have found that the edge states are ubiquitously hybridizing
with the spin-split electronic states in the FM. 
As a direct consequence, the spectral weight of the midgap states is reduced and the spin-current along the interface is highly suppressed. 
More specifically, spin-up edge states keeps the same degree of hybridization for about $h<1.0 t$ and thus the spin up current $J^{tot}_{\ua}$
stays essentially unchanged. 
On the contrary, for the spin-down channel, one can observe that the states with momenta around $k_y=\pm \pi/2$ 
are less and less modified when $h$ is increasing and, therefore, $J^{tot}_{\da}$ gets more electronic contributions and 
tends to grow.

Differently from the regime of weak-intermediate FM, above an exchange threshold of the order of $h=1.0 t$, both $J^{tot}_{\ua}$ and $J^{tot}_{\da}$ exhibit a monotonous upturn with 
an almost linear trend. This dependence on the exchange field in the regime of a large FM magnetization can be justified by
noticing that the spin-up polarized states with momentum around $k_y=0$ and those with spin-down configuration 
close to $k_y=\pm \pi/2$ become less hybridized with the ferromagnet spectrum. 
Therefore, the spin currents carried by the unmodified part of the edge states 
are increasing for both spin polarization.\\

Taking into account the different regimes of the integrated spin-current for the case of \None, it is useful to 
consider how the capacity of carrying spin-currents is modified by the presence of additional helical modes. The first difference to point out is that $J^{tot}_s$ is more suppressed, with respect to the vacuum-TTSC interface amplitude, for the case \None than for \Ntwo.
This quantitative disparity arises from the fact that
the \Ntwo edge states 
remain not hybridized close to the zone boundary at $k_y=\pm \pi$ and, thus, still have
a net capacity to carry spin-current without the effect of the mixing with the 
metallic electronic states close to the Fermi level.
The second relevant difference between the \None and \Ntwo helical superconductor is
observed in the regime of weak-to-intermediate ferromagnet. 
In this range, while $J^{tot}_{\ua}$ remains constant when $h$ is growing from $h=0$ to $1.0 t$ for the case \None,
the \Ntwo current response exhibits a monotonous decrease. As we can observe in the spectral functions of Fig.~\ref{fig.spectrum.spin.up}, 
this is directly related to the evolution of the majority spin-polarized edge states with momentum close to 
$k_y= \pm \pi$ where the degree of hybridization 
with the ferromagnetic states is more pronounced and increases with the exchange field. Hence, 
the capacity of carrying current along the interface results into a net suppression 
and the total spin current $J^{tot}_s$ 
remains constant in this range of values for the exchange field.
It is worth pointing out that the amplitude of the spin-polarized currents is not only due to the midgap helical edge states at the boundary of the TTSC but there
is also a tiny contribution arising from the states above the superconducting gap. Since this part is negligible with respect to that of the midgap states 
the current variations with respect to the exchange field can be addressed by
focusing only on the part of edge-states spectral functions as in Figs.~\ref{fig.spectrum.spin.up} and~\ref{fig.spectrum.spin.down}.

\begin{figure}[!t]
\begin{center}
\includegraphics[width=0.43\textwidth]{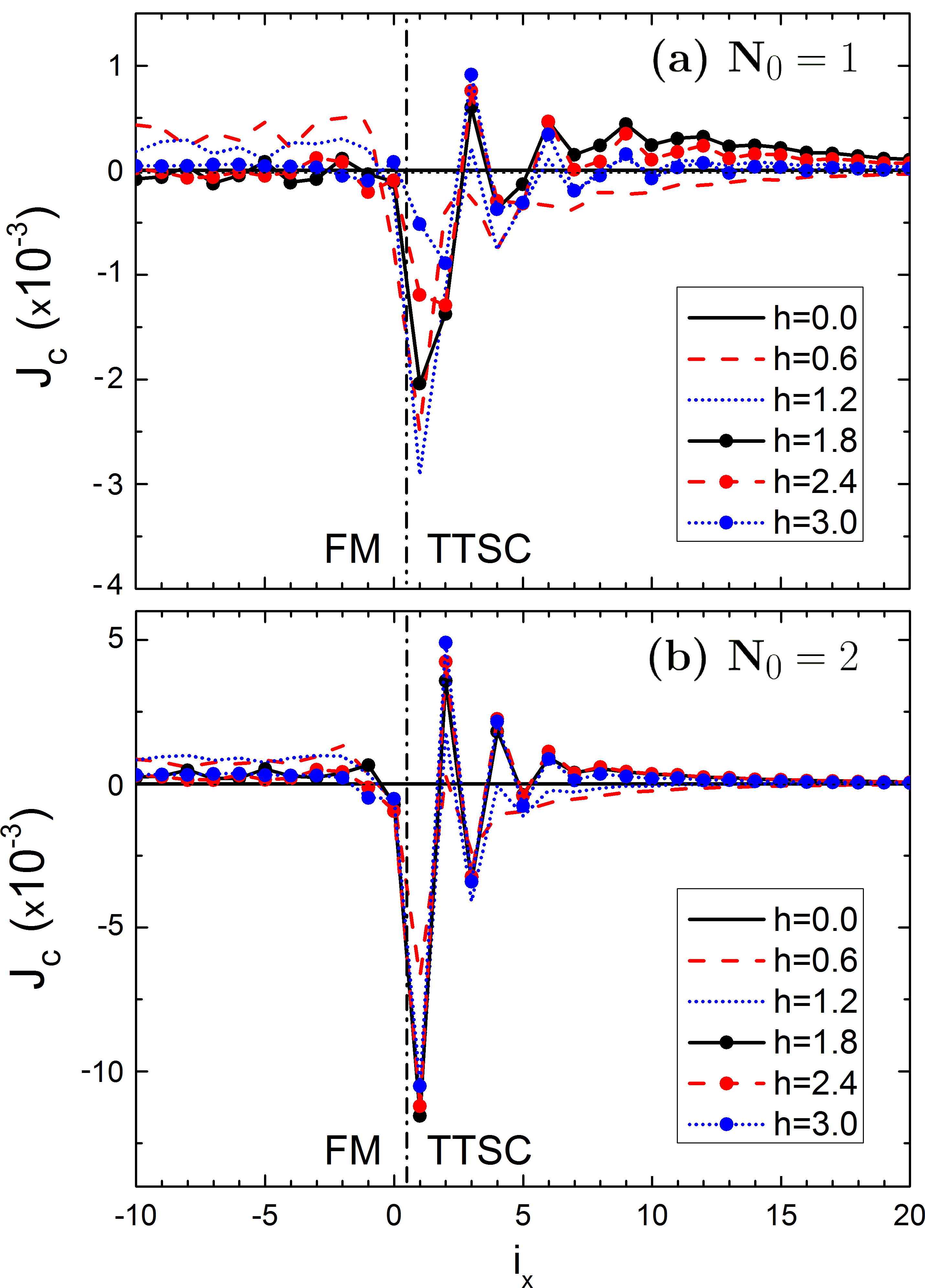}
\end{center}
\caption{Spatial variation of the charge current $J_c(i)$ along the FM-TTSC interface for topological configurations with $N_0=1$ (a) and $N_0=2$ (b) helical modes. 
The magnetization is perpendicular to the $\vec{d}$, i.e. $\theta=0$. The dashed-dotted line represents the interface between the FM and TTSC regions.}
\label{fig.Jcharge.t0}
\end{figure}

As discussed previously, the FM breaks the up-down spin symmetry and, thus, leads to a net integrated charge currents due to the 
difference between the majority and minority spin electrons currents.  
While within the TTSC there is a definite relation between the integrated spin-up and
spin-down currents with a general tendency to provide a 
negative total charge current being $-J^{tot}_{\ua}< J^{tot}_{\da}$, the spatial
dependence of the charge current is non-monotonous and exhibits a decaying 
behavior moving towards the inner side of the TTSC and the FM with an oscillatory component whose amplitude
scales with the exchange field. This is shown in Fig.~\ref{fig.Jcharge.t0} where we present the site dependent evolution of the charge current $J_c(i_x) = J_{\ua}(i_x) + J_{\da}(i_x)$ in the proximity of the FM-TTSC interface, for both \None(a) and \Ntwo(b) superconductors, for small to large values of the exchange field $h$. 

The spatial behavior of the charge current reveals a subtle dependence on the hybridization between the ferromagnetic and the helical states 
that leads to a distinct response at the two sides of the heterostructure and is qualitatively sensitive to the strength 
of the FM exchange.
As a general trend, we can note that the charge current is maximal at the interface and then decreases when one moves away from the interface. 

A peculiar aspect is represented by the flow direction of the charge current in the TTSC and FM domains. 
For weak-to-intermediate FM, the charge current has a one-way flow within the TTSC and the FM side of the heterostructure with 
opposite relative direction. Moreover, for such a range of exchange fields 
the size of the charge current grows (decreases) in the TTSC (FM) regions, respectively.
When the exchange in the FM overcomes a critical threshold of about $h\sim 1.0 t$ with a 
magnetization that is larger than half of its maximal value, the behavior 
of the charge currents is completely modified as it becomes oscillatory 
with a change in the flow direction that depends on the distance from the interface.  
An important feature is that the oscillatory length scale does not depend on the strength of the exchange $h$, being hence uncorrelated to the spin split of the magnetic states in the FM but connected to 
the spatial dependence of the spectral weight renormalization of the helical states. These characteristics are fundamentally common to the \None and \Ntwo TTSC.

Among the \None and \Ntwo differences we point out that they emerge mainly in the
strong FM regime where the charge currents oscillates with a sign change 
in the ferromagnetic region for \None
while it keeps the same sign (flow) for the case of \Ntwo TTSC.
Another distinct aspect is that the amplitude of the charge current 
at the interface within the TTSC domain is more sensitive to the exchange for the \None 
than the \Ntwo helical superconductor.\\

\begin{figure}[!t]
\begin{center}
\includegraphics[width=0.43\textwidth]{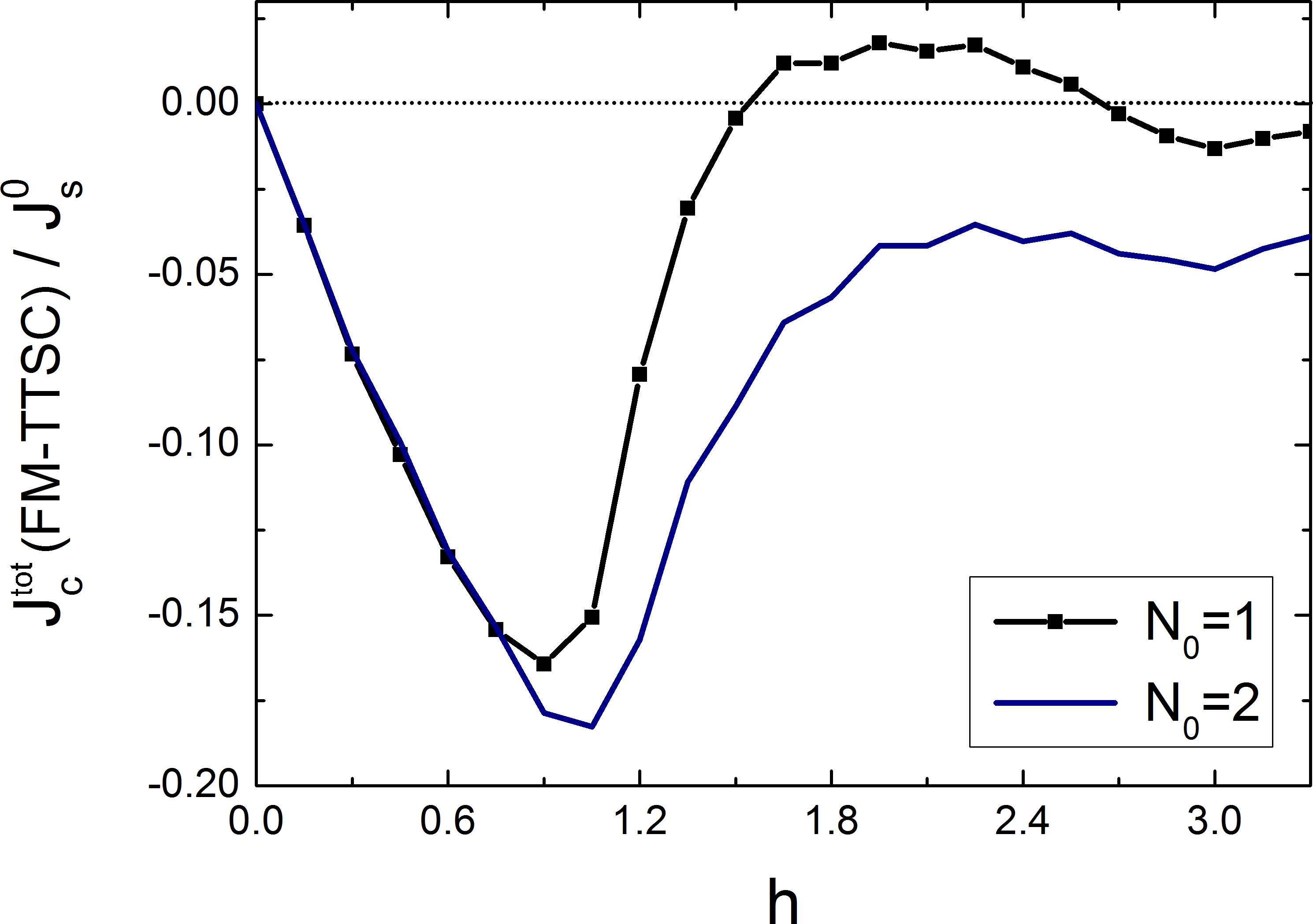}
\end{center}
\caption{Integrated charge current inside the TTSC region for various strengths of the exchange field $h$ and normalized to its value for the vaccum-TTSC interface. 
The results are showed for configurations with $N_0=1$ (a) and $N_0=2$ (b) helical modes. 
They are obtained from $J_{\uparrow}$ and $J_{\downarrow}$ of the Fig.(?).
The magnetization is assumed perpendicular to the d-vector, i.e. $\theta=0$.}
\label{fig.Jcharge.t0.sum}
\end{figure}

For further pinpointing the differences between the charge currents response for the \None and 
\Ntwo helical superconductors, it is useful to investigate the integrated amplitude within the TTSC 
region of the heterostructure.
As we have done previously for the spin currents, we evaluate the total charge current $J^{tot}_c = J^{tot}_{\ua} + J^{tot}_{\da}$ 
flowing in the TTSC region and compare its evolution as a function of the exchange field in the FM for the \None and \Ntwo cases (Fig. \ref{fig.Jcharge.t0.sum}). 
The spin-polarized currents have been already shown in Fig.~\ref{fig.Jspin.t0.sum}.

Here, due to the presence of the spatial oscillatory behavior, the results of $J^{tot}_c$ cannot be easily inferred from the
variation of $J_c(i_x)$. Indeed, for small values of the exchange field, i.e. in the range $h=0$ to $h=1.0 t$, 
it is interesting to see that the charge current is insensitive to the number of helical modes. $J^{tot}_c$ reaches its maximum 
at about $h=1.0 t$ where the ferromagnet magnetization approaches about the half of the fully polarized configuration. 
Then, due to the reduction of the mixing between the ferromagnetic and the helical modes, 
$J^{tot}_c$ decreases when the exchange field moves towards the strong ferromagnet limit.

In this range we can observe the main difference between the two helical cases, i.e. for $h>1.5 t$. While the charge current remains 
non zero and almost constant for \Ntwo when the magnetization gets to the half-metallic limit, 
for the \None case it gets smaller with a sign change accompanied by an oscillatory behavior  
for large exchanges $h>1.5 t$. This qualitative difference can be understood by looking at the spectral functions for spin-up 
and spin-down electrons, Figs.~\ref{fig.spectrum.spin.up} and ~\ref{fig.spectrum.spin.down}. Indeed, by approaching the half-metallic 
regime for \None, the Fermi surfaces of the spin-up and spin-down electrons are such that the edge states, with momentum $-\pi/2<k_y<\pi/2$, 
are mostly not hybridizing with the ferromagnetic spectra. Hence, $J^{tot}_{\ua}$ and $J^{tot}_{\da}$ are close 
to the expected values for the the vacuum-TTSC interface configuration. However, such a decoupling does not occur for \Ntwo. 
The majority spin edge 
states close to $k_y=\pm \pi$ are mixed with the ferromagnetic electronic states close to the Fermi level. 
Therefore, $J^{tot}_{\ua}$ tends to 
diminished and there is no compensation in amplitude between the opposite spin-polarized currents with a resulting
net charge current flowing at the interface.

These results for FM magnetization that is perpendicular to the plane indicates that the evolution of the integrated charge current at the interface with respect 
to the strength of the exchange field can give important hints to discern between superconductors having different
number of helical modes at the edge.

\subsection{Magnetization coplanar to the $\vec{d}$-vector}

The spin and charge currents at the FM-TTSC interface exhibit a behavior that 
includes new effects when the magnetization is coplanar to the 
spin-triplet $\vec{d}$-vector order parameter. 
Since the components of the $\vec{d}$-vector 
have different orbital character with respect to
the interface orientation, the spin content of the helical states
depends on the transverse momentum along the interface. Hence, the
coupling of the ferromagnetic spin-exchange to the helical states with
$x$ and $y$ zero spin-projections
naturally leads to an anisotropic response and to a
momentum-dependent modification of the electronic spectrum.
We remind that the physical configuration of the FM-TTSC heterostructure analyzed here 
has an interface orientation that is
perpendicular to the $x$ component of the $\vec{d}$-vector with a $p_x$ orbital 
symmetry.

\begin{figure}[!b]
\begin{center}
\includegraphics[width=0.43\textwidth]{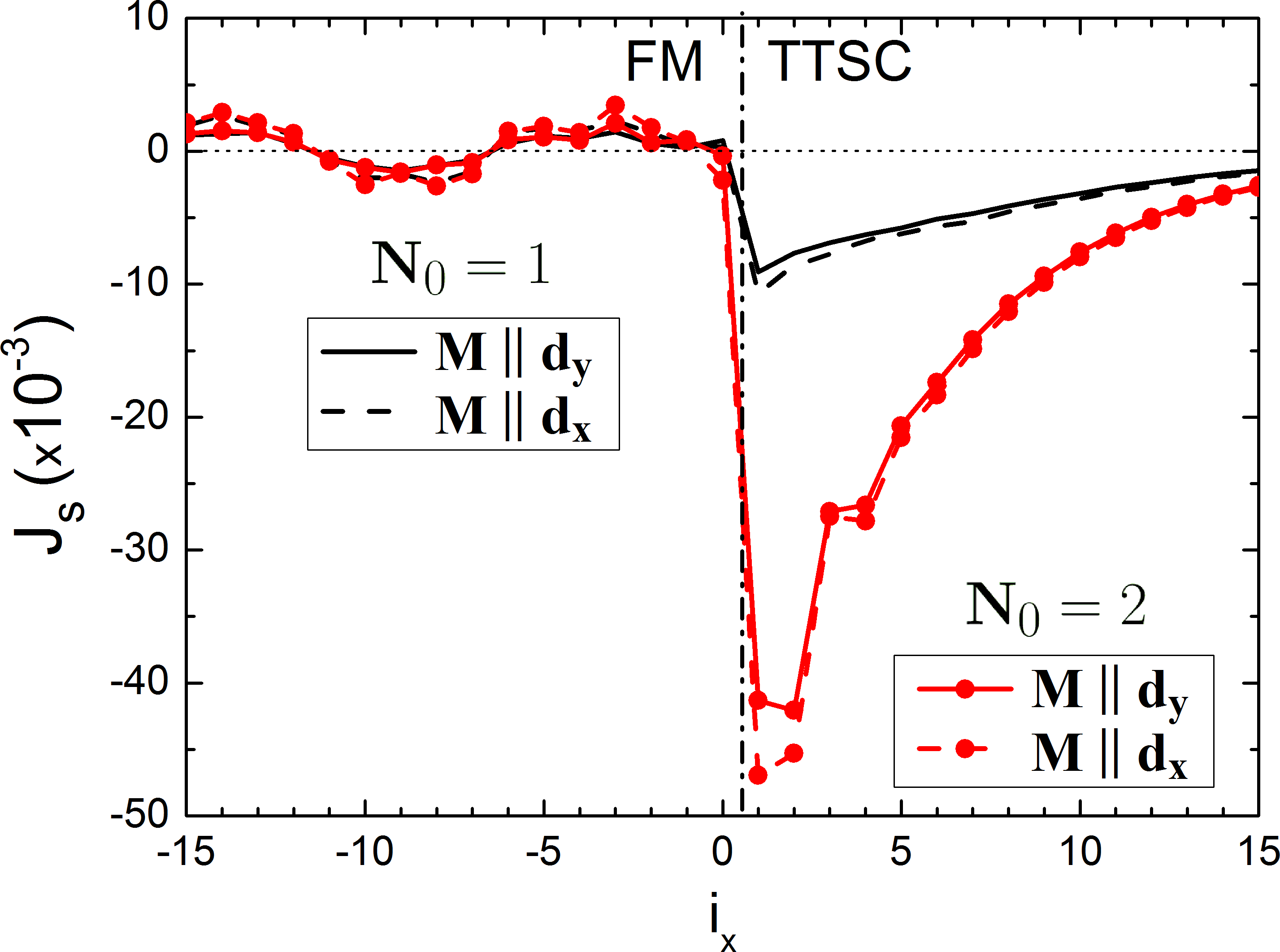}
\end{center}
\caption{Evolution of the spatial variation of the $z$-polarized spin current $J^{z}_s(i_x)$ along the FM-TTSC interface at a representative value of the exchange $h=1.0 t$. 
The plot includes the behavior for both \None (black) and \Ntwo (red) TTSC. 
The magnetization is oriented in-plane along the $y$-direction (solid lines) and the $x$-direction (dashed lines). 
The dashed-dotted line represents the interface between the FM and TTSC regions.}
\label{fig.Jspin.t90}
\end{figure}

The first distinctive feature of a ferromagnetic exchange that is coplanar to the 
$\vec{d}$-vector is provided by the up-down symmetry of the electronic spectrum. 
Hence, the relation $A_{\ua}(i_x,k_y)(\omega)=A_{\da}(i_x,-k_y)(\omega)$ hold at any site of the system and for all strength of the exchange field $h$. It leads to a spin resolved current $J_{\ua}(i_x)=-J_{\da}(i_x)$ and, therefore, a vanishing
charge current at the interface, $J_c(i_x)=0$. We present in Fig.~\ref{fig.Jspin.t90} the comparison between the spatial variation of the $z$-polarized spin current for 
two orientations of the exchange and a representative value of $h=0.6t$  for \None and \Ntwo TTSC. 
We can observe that, due to the proximity effect, the spin current is oscillating in the ferromagnetic region. Indeed, 
an in-plane magnetization acts as a pair-breaking of the Cooper pairs that penetrate into the FM and in turn induces oscillations 
in the spin current. Due to the $xy$ symmetry of the pair-breaking mechanism, 
the period of the oscillations does not depend on the orientation of the in-plane magnetization and it is 
weakly linked to the number of helical modes. When $h$ is increasing, the period of the oscillations shrinks and the 
amplitude is suppressed, such that there is no induced spin current in the ferromagnetic region for the regimes of  intermediate-strong FM. 

On the other hand, the spatial variation of $J^{z}_s$ in the TTSC is very similar to the case at $\theta=0$ with a magnetization that is
perpendicular to the $\vec{d}$-vector. Indeed,  $J^{z}_s$ reaches its  maximal value at the the boundary of the TTSC while
it decreases over a distance $\xi_{sc} = 20$ sites in the TTSC side. 
Additionally, for both cases \None and \Ntwo, we can see that the spin current along the interface is slightly larger when 
the magnetization is parallel to $d_x$ than to $d_y$. This effect is due to the gap opening at zero momentum for the Majorana states 
associated with the $p_x$ orbital symmetry of the $d_y$ component of the spin-triplet order parameter (Fig.~\ref{fig.spectrum.spin.down.t.90}). 
The anisotropy is small for the weak FM regime because the edge states are completely mixed close to $k_y=0$ and hence 
the tendency to split of the Majorana modes cannot fully contribute to the decrease of the spin current.\\

\begin{figure}[!t]
\begin{center}
\includegraphics[width=0.43\textwidth]{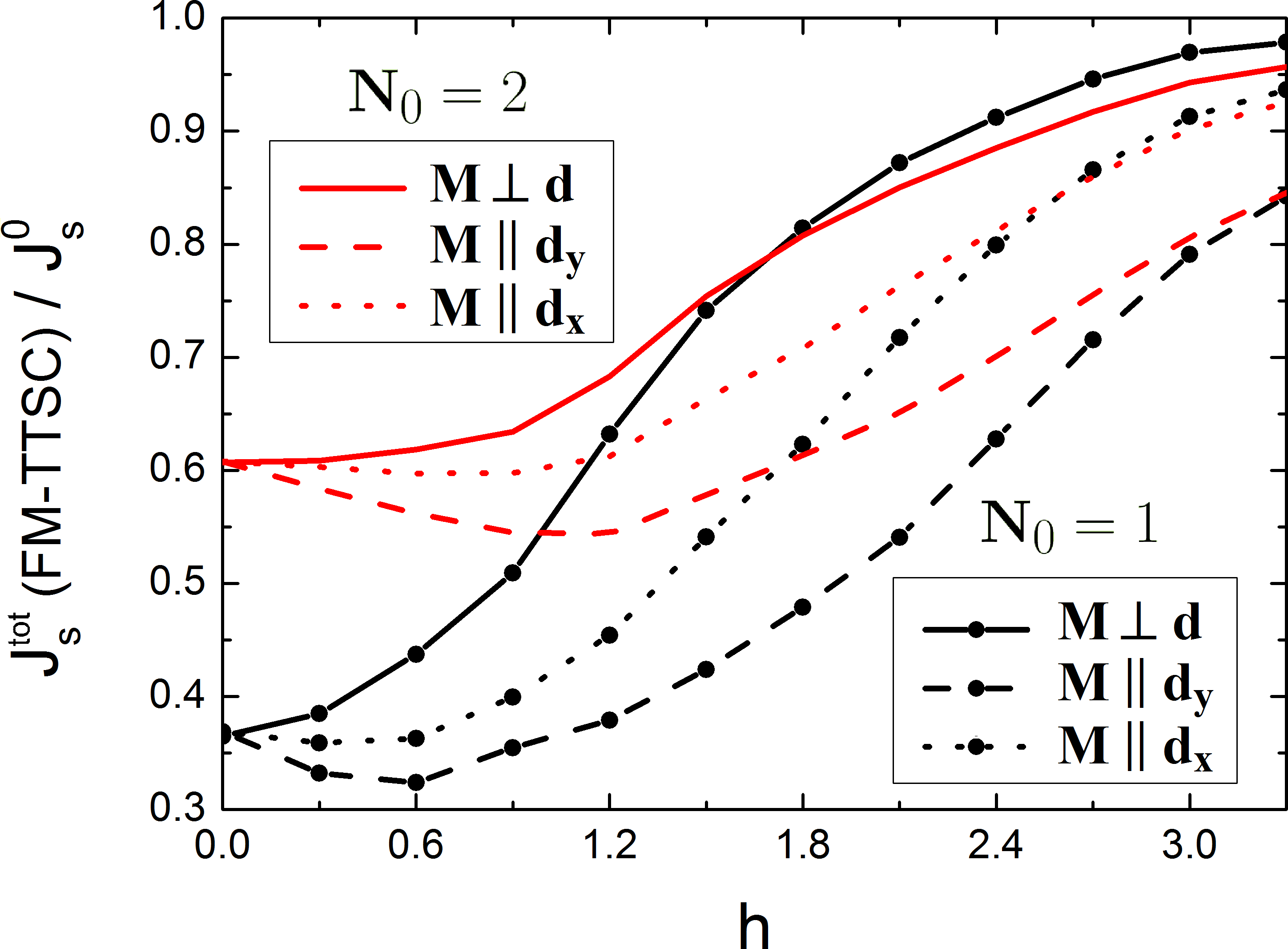}
\end{center}
\caption{(color online). Evolution of the integrated spin current $J^{tot}_s$ in the TTSC region, as 
normalized to the value at the vaccum-TTSC interface, versus the exchange field $h$. 
The results are presented for the $N_0=1$ (black) and $N_0=2$ (red) cases with a magnetization 
in the FM oriented along the $d_z$- (solid), $d_y$- (dashed) and $d_x$ directions (dotted line), 
respectively.}
\label{fig.Jspin.t90.sum}
\end{figure}

The in-plane orientation's dependence of the spin current for coplanar magnetization is significantly 
linked to the intensity of the FM and to the number of helical modes in the TTSC. In order to emphasize this point, we present in Fig.~\ref{fig.Jspin.t90.sum} the evolution of the $z$-polarized spin current $J^{tot}_Z$ integrated within the TTSC region as a function of the exchange field $h$, for several orientation of the magnetization in the ferromagnet. As a general observation, we can distinguish three different regimes for the spin-current response when the magnetic exchange is tuned from weak to strong FM.

For the case of weak FM and a magnetization that is longitudinal to the $d_x$ component is basically 
flat independently of the number of helical modes. On the other hand, 
an orientation that is parallel to $d_y$ generally leads to 
a decrease of the spin-current. The increase in the exchange amplitude leads to an upturn 
of the spin-current at lower values of the magnetization for the case \None than for \Ntwo TTSC.

The particularity of the \Ntwo TTSC is that, to achieve a spin-current that is
larger than the value of the normal-TTSC interface, one needs to 
approach the regime of intermediate-strong FM at any 
orientation of the magnetization coplanar to the $\vec{d}$-vector.
On the contrary, while approaching the half-metallic FM configuration,
the spin-current response of the FM-TTSC does not exhibit 
significant differences between the two case, both qualitatively and quantitatively.

\begin{figure}[t!]
\begin{center}
\includegraphics[width=0.43\textwidth]{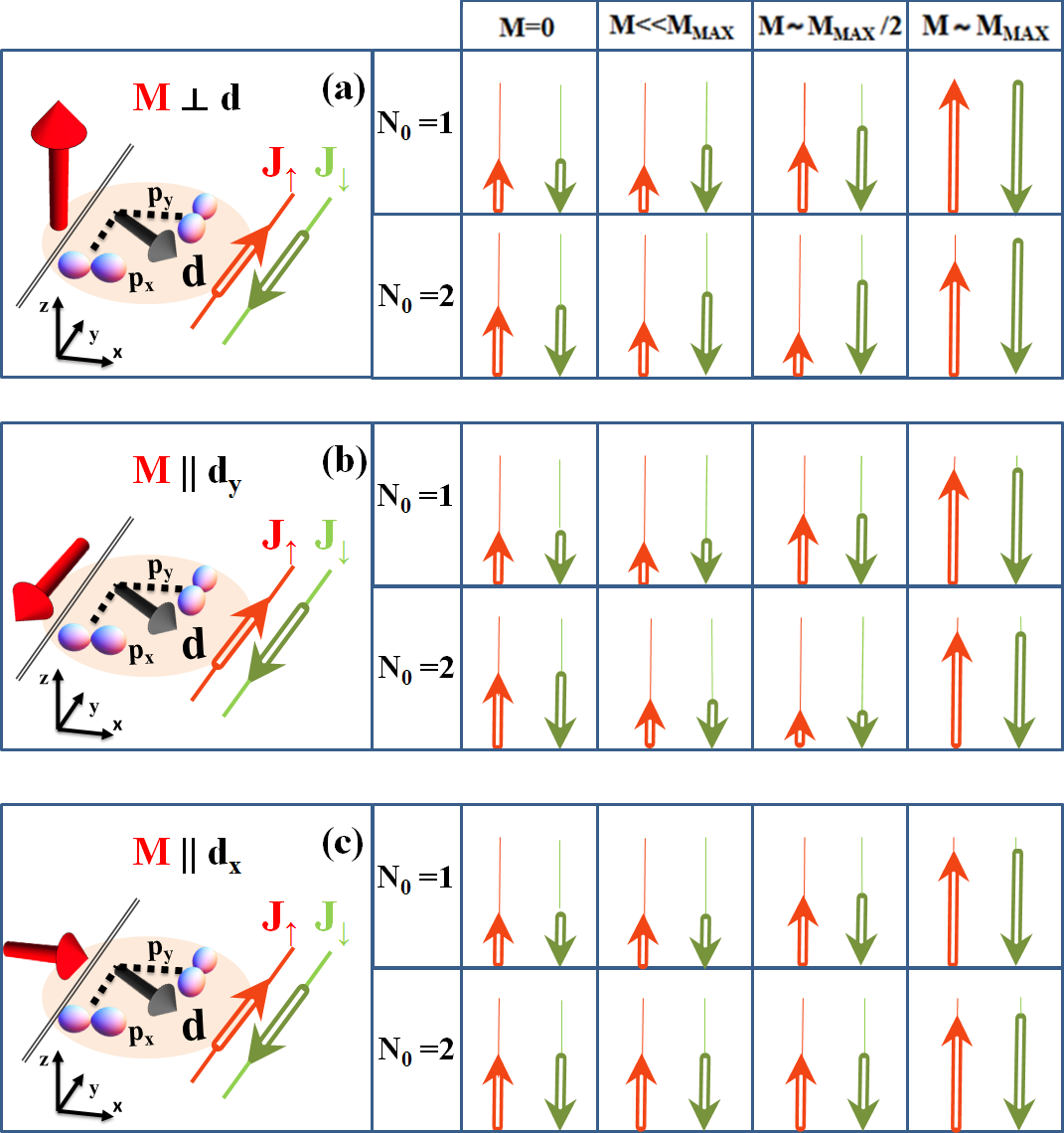}
\end{center}
\caption{(color online). Conclusive sketch illustrating the strengths of the spin-up (red arrows) and spin-down (green arrows) currents at the N-TTSC ($M=0$) and FM-TTSC interfaces, for weak ($M \ll M_{max}$), intermediate ($M \sim M_{max}/2$) and strong ($M \sim M_{max}$) regimes. They are shown for different orientations of the ferromagnetic magnetization and number $N_0$ of helical modes. The thin lines depict the maximal strength of the spin-polarized currents at the vacuum-TSC interfaces. Finite spin currents are present for each configurations and charge currents are induced only in the configuration (a) when $J_{\ua} \neq - J_{\da}$.}
\label{fig.Jspin.t90.sum.sketch}
\end{figure}


\section{Conclusions}

We have studied the spin and charge current behavior of an FM-TTSC heterostructure as a function of 
the orientation and amplitude of the magnetization in the ferromagnet and for superconductors having one and two pairs of helical modes at their edge. 
Our main results concern the capacity of tuning the spin and charge currents as well as the individual spin polarized component 
$J_{\sigma}$ 
by means of the number of helical modes in the spin-triplet superconductor and the 
amplitude-orientation of the ferromagnetic exchange with respect to the ${\vec{d}}$ vector. 
The presence of two helical channels allows 
to have opposite trends in the current response of the majority and minority spin
channels that are quantitatively and qualitatively different with respect to the single pairs of 
helical modes, as sketched in Fig. 13, in all regimes from weak-to-strong
ferromagnet. The anisotropic hybridization of the FM states with the helical modes 
is shown to be the key dynamical parameter that controls the current behavior.
While our study has been performed for a specific shape of the FM Fermi surface,
the unveiled mechanisms allow to immediately generalize the current response to a suitably
designed FM material with single or multiple bands hybridizing with the helical modes.  

Due to the symmetry of the superconductor, we find that the current polarization is mainly perpendicular 
to the ${\vec{d}}$-vector, and two main 
behaviors can be achieved with flat and monotonous trend depending on the strength of the FM magnetization. 
The presence of multiple helical modes at the edge makes generally
more difficult to recover the capacity of carrying spin current with an amplitude that is comparable 
to that observed at the edge of the TTSC with the vacuum. The two-mode helical TTSC 
can exhibit an almost flat behavior with a magnetization 
being not longitudinal to the $\vec{d}$-vector component and an 
orbital symmetry that is perpendicular to the 
interface (i.e. $d_y$ for the analyzed configuration).  
For a magnetization that is
transversal to the $\vec{d}$-vector, the flatness is
generated by a peculiar compensation of the opposite
spin-polarized currents.
On the contrary, an exchange that is able to 
split the time-reversal Kramers pairs tends 
to open a gap in the helical electronic spectra and 
results into a suppression of the spin-current.
The $N_0=2$ TTSC tends to amplify this 
effect and the spin-current keeps decreasing for a large range
of exchange fields.

The orientation dependence underlines a strong difference when 
the FM-TTSC heterostructure has an FM magnetization that is perpendicular to the
$\vec{d}$-vector exhibiting a nonvanishing total charge current that can flow 
together with the spin current. 
We find that the charge current amplitude is maximal 
for intermediate strength of the FM and exhibits 
the same response at weak FM independently of the number of helical modes.
While the integrated current has a net flow within the TTSC region close 
to the interface, the spatial dependence of the charge current
shows a complex sign changing profile that 
changes with the distance from the interface.
The spatial dependence of the current reveals 
the subtle interplay behind the mixing of the
magnetic and helical states at the interface.
Finally, the capacity of carrying spin current in the FM 
depends on the orientation of the magnetization and 
can be even counterpropagating with respect to that 
flowing in the TTSC.


\acknowledgments

The authors thank A.~Schnyder for useful discussions and G.~Khaliullin for his careful reading of the manuscript. D.~T. gratefully acknowledges the hospitality of the University of Salerno and CNR-SPIN where part of this work was done.


\end{document}